\documentclass[a4paper,11pt]{article}
\usepackage{pos}
\usepackage{amsmath,graphicx,multirow,tabularx,physics}
\usepackage{davitex,slashed,lineno}

\newcommand{\Pbare}{P^{\text{bare}}}

\newcommand{\freg}{f_{\text{reg}}}
\newcommand{\fresponse}{\chi_{Q,\mu_B^2}}
\newcommand{\muh}{\hat{\mu}}

\title{Imprint of chiral symmetry restoration on the Polyakov loop and the heavy
quark free energy}
\ShortTitle{Chiral Imprint on Polyakov Loop}

\author*[a]{D. A. Clarke}
\author[a]{O. Kaczmarek}
\author[a]{F. Karsch}
\author[a]{Anirban Lahiri}
\author[a,b]{Mugdha Sarkar}

\affiliation[a]{Fakult\"at f\"ur Physik, Universit\"at Bielefeld,\\
 Bielefeld, Germany} 
\affiliation[b]{Department of Physics, National Taiwan University, \\Taipei 10617, Taiwan} 

\emailAdd{dclarke@physik.uni-bielefeld.de}
\emailAdd{okacz@physik.uni-bielefeld.de}
\emailAdd{karsch@physik.uni-bielefeld.de}
\emailAdd{alahiri@physik.uni-bielefeld.de}
\emailAdd{mugdha@physik.uni-bielefeld.de}

\abstract{The Polyakov loop expectation value $\ev{P}$ is an order parameter of 
the deconfinement transition in the heavy quark mass regime, whereas its sensitivity 
to the deconfinement of light, dynamical quarks is not apparent. From the perspective 
of an effective Lagrangian in the vicinity of the chiral transition, the Polyakov 
loop, $P$, is an energy-like observable, and $\ev{P}$ should hence scale like the 
energy density. Using $N_f=2+1$ HISQ configurations at finite lattice spacing, we 
show that near the chiral transition temperature, the scaling behavior of
$\ev{P}$ and the heavy quark free energy $F_q$ is consistent with energy-like 
observables in the 3-$d$, $\O(N)$ universality class. We extend this analysis to 
other Polyakov loop observables, including the response of the heavy 
quark free energy, $F_q$, to the baryon 
chemical potential, which is expected to scale like a specific heat.}

\FullConference{%
 The 38th International Symposium on Lattice Field Theory, LATTICE2021
  26th-30th July, 2021
  Zoom/Gather@Massachusetts Institute of Technology
}

\begin{document}
\maketitle

%
%
\section{Introduction}

The Polyakov loop and its spatial average are defined in lattice QCD by
\begin{linenomath*}\begin{equation}
  P_{\vec{x}}\equiv\frac{1}{3}\tr \prod_\tau U_4\left(\vec{x},\tau\right),
  ~~~\text{and}~~~
  P\equiv\frac{1}{N_\sigma^3}
  \sum_{\vec{x}}P_{\vec{x}},
\end{equation}\end{linenomath*}
where the space-time volume is $N_\sigma^3\times N_\tau$ and
$U_4(\vec{x},\tau)\in\SU(3)$ is the link variable
originating at the site $(\vec{x},\tau)$ and pointing in the
Euclidean time direction. $P$ can be related to the heavy-quark free
energy by\footnote{We have made explicit in this equation the dependence of 
$P$ and $F_q$ on the temperature $T$ and $H$. 
We will sometimes not explicitly write these
dependencies to keep the notation light.}
\begin{linenomath*}\begin{equation}\label{eq:Fav}
  F_q(T,H)  = -T\ln \ev{P(T,H)} 
  = -\frac{T}{2} \lim_{\left|\vec{x}-\vec{y}\tinysp\right|\rightarrow \infty}
\ln \ev{P^{\phantom\dagger}_{\vec{x}} P^\dagger_{\vec{y}}},
\end{equation}\end{linenomath*}
where $T$ denotes the temperature and $H=m_l/m_s$ is the ratio of the light and
strange quark masses $m_l$ and $m_s$.
In the quenched limit, $\ev{P}$ is an order parameter signaling the
deconfinement of static quarks above a critical temperature $T_d$, where a global
$\Z_3$ symmetry is spontaneously broken. 

At finite quark mass this symmetry is broken explicitly, and it is unclear 
whether it makes sense to associate
large changes in $\ev{P}$ with the deconfinement of light quarks in particular.
Still, there is a clear $T$-dependence for $\ev{P}$ at all quark masses,
which is often viewed as a remnant of $m_l=\infty$ physics, including its interpretation
in terms of deconfinement. This viewpoint is perhaps motivated in part by a 
coincidence of inflection points for $\ev{P}$ and the chiral condensate
in the quenched limit~\cite{kogut_deconfinement_1983} and a seeming
coincidence at larger-than-physical light quark masses~\cite{cheng_qcd_2008},
although the latter coincidence disappears closer to the continuum
limit~\cite{Aoki:2009sc,Bazavov:2013yv,Bazavov:2016uvm,Clarke:2019tzf}.

One aim of this project therefore was to understand how to interpret the 
$T$-dependence of $\ev{P}$. There has been much effort investigating the order
of the chiral phase transition, and in this context it is natural to examine
to what extent the behavior of $\ev{P}$ and related observables such as $F_q$ 
is determined by chiral scaling. For example the response of 
$F_q$ to the baryon
chemical potential $\mu_B$, 
\begin{equation}\label{eq:fresponse}
 \fresponse \equiv - \frac{\partial^2F_q/T}{\partial\muh_B^2}\,\Big|_{\muh_B=0},
 ~~~~~\hat{\mu}_B\equiv\mu_B/T
\end{equation}
shows, for larger-than-physical
pion mass using the p4-action~\cite{Doring:2005ih} and recently
for HISQ quarks at physical mass~\cite{delia_dependence_2019},
a peak near the QCD crossover. It is similarly interesting
to see whether the peak in this observable, which is also derived from
the Polyakov loop, can be understood through chiral scaling as well.

In these proceedings we summarize our findings
in Ref.~\cite{Clarke:2020htu}, that at physical and smaller $m_l$, 
the Polyakov loop and related observables such as $F_q$ are influenced by the 
chiral transition. Data used for this part of the project that are
presented in the figures can be found online in Ref.~\cite{Clarke:2020data}.
We also report our progress analyzing $\fresponse$.

%
%
\section{Chiral scaling of Polyakov loop observables}

QCD thermodynamics in a neighborhood of the chiral critical point
at $m_l=0$ and critical temperature $T_c$ can be
described by some effective Lagrangian. Any operator appearing in this
Lagrangian may either break the global $\SU(2)_L\times\SU(2)_R$ chiral 
symmetry explicitly or respect it. In the former case, the operator comes with a
symmetry-breaking coupling, and we call the operator 
magnetization-like\footnote{We are borrowing nomenclature from spin systems.}; 
in the latter case it is an energy-like operator. Since $P$ is trivially
invariant under chiral rotations, it is an energy-like operator, {\it i.e.}
it has the same scaling function as the energy density.
Near the critical point, energy-like observables can be
expressed as the sum of a regular part, which is an analytic function
in the symmetry-breaking parameter $H$ and the reduced temperature
$t\equiv(T-T_c)/T_c$, and a singular part, which is determined by 
a universal scaling function of the scaling variable 
$z\equiv z_0 t H^{-1/\beta\delta}$. Here $\beta$ and $\delta$ are universal 
critical exponents, and $z_0$ is a non-universal constant that sets the scale of $z$.

In the continuum limit with two light flavors, the chiral phase transition is expected 
to fall in the $\O(4)$ universality
class~\cite{pisarski_remarks_1984}, and while this is not known with 
absolute certainty, recent lattice calculations present evidence favoring this 
scenario~\cite{HotQCD:2019xnw,Cuteri:2021ikv}. Therefore in this study we used scaling
functions belonging to the 3-$d$, $\O(N)$ universality class.
We used HISQ fermions without performing a continuum limit
extrapolation; therefore due to taste violations, the relevant universality
class is 3-$d$, $\O(2)$ with critical exponents\footnote{The $\O(2)$
critical exponents $\beta$ and $\delta$ are close to those of $\O(4)$, and 
importantly $\alpha<0$ for both universality classes.}~\cite{Engels:2000xw}
\begin{linenomath*}\begin{equation}
\beta=0.349, ~~~~\delta=4.780, ~~~~\text{and}~~~~
\alpha=2-\beta(1+\delta)=-0.0172.
\end{equation}\end{linenomath*}

Since $F_q/T$ is an energy-like observable, its scaling behavior can be
expressed as~\cite{Engels:2011km}
\begin{linenomath*}\begin{equation}
                F_q/T = A H^{(1-\alpha)/\beta\delta} f'_f(z) 
                +f_{\rm reg}(T,H),
\label{Fqcritical}
\end{equation}\end{linenomath*}
where $A$ is another non-universal constant,
$f'_f(z)$ is the $z$-derivative of the scaling
function $f_f$ characterizing the singular part of the logarithm of
the partition function, and
\begin{linenomath*}\begin{equation}
                \freg= \sum_{i,j}  a^r_{i,2j}\ t^i H^{2j} 
                \equiv \sum_j p^r_{2j}(T) H^{2j}.
\label{freg}
\end{equation}\end{linenomath*}
Note that we consider above the infinite volume scaling ansatz since we perform 
our analysis on data obtained from simulations with the largest volumes.
The strategy is to start with the scaling behavior\footnote{In principle one 
could have taken $\ev{P}$ as the starting point.} eq.~\eqref{Fqcritical} and
derive the other observables from this.
For instance from eq.~\eqref{Fqcritical} and eq.~\eqref{eq:Fav} one finds 
\begin{linenomath*}\begin{equation}
                \ev{P} = \exp\left(-
                A H^{(1-\alpha)/\beta\delta} f'_f(z) 
                -f_{\rm reg} \right).
                \label{Pcritical}
\end{equation}\end{linenomath*}
Also making use of the relation between $f_f$ and the order parameter scaling
function $f_G$,
\begin{linenomath*}\begin{equation}\label{eq:fGandff}
f_G(z) = -\left(1+\frac{1}{\delta}\right) f_f(z)
+\frac{z}{\beta\delta}f'_f(z)~,
\end{equation}\end{linenomath*}
we obtain the quark mass dependence near the critical point
\begin{linenomath*}\begin{equation}
\frac{\partial F_q/T}{\partial H} = -A H^{(\beta -1)/\beta\delta}f'_G(z) 
             +\frac{\partial f_{\rm reg}}{\partial H} .
\label{Fqm-crit}
\end{equation}\end{linenomath*}
Equation~\eqref{Fqm-crit} is especially interesting for us; since
$(\beta-1)/\beta\delta<0$, this quantity will diverge in the chiral limit, meaning that
the singular contribution will dominate for sufficiently small $H$.
Finally for the temperature dependence, we find 
\begin{linenomath*}\begin{equation}\label{eq:FqT}
T_c\frac{\partial F_q/T}{\partial T} =  
A z_0 H^{-\alpha / \beta\delta} f''_f(z) + T_c \frac{\partial f_{\rm reg}}{\partial T}  .
\end{equation}\end{linenomath*}

In order to carry out fits to the above functional forms, we need some detailed
information about the scaling function $f_f$. We do this using expansions
of $f_f$ about $z=0$ and $z=\pm \infty$, using the notation of 
Ref.~\cite{Engels:2011km} and fitting to the 3-$d$, $\O(2)$ data of 
Ref.~\cite{Engels:2000xw}.
Inserting this expansion into eq.~\eqref{Fqcritical} we 
obtain at fixed $T$ for small $H$
\begin{linenomath*}\begin{equation}
		\frac{F_q}{T} \sim 
\begin{cases}
a^-(T)+A p_s^-(T)\ H &,\ T< T_c \\
a^r_{0,0}+A a_1\ H^{(1-\alpha)/\beta\delta} &,\ T=T_c \\
a^+(T)+p^+(T)\ H^2 &,\ T> T_c
\end{cases}
\; ,
\label{O4criticalmass}
\end{equation}\end{linenomath*}
where $a^\pm(T)=A a_s^\pm(T)+\freg(T,0)$ and $p^+(T)=A p_s^+(T)+p^r_2(T)$ 
receive contributions from both singular and regular terms. 
Below the critical temperature, the 
dominant quark mass dependence arises from the singular 
term only\footnote{This linear term in $H$ is consistent with what one
expects from a heavy-light resonance 
gas~\cite{Megias:2012kb,Bazavov:2013yv} 
and heavy-quark effective theory/chiral perturbation theory~\cite{Brambilla:2017hcq}.
For more details see Ref.~\cite{Clarke:2020clx}.}. We find 
\begin{linenomath*}\begin{eqnarray}
\label{coefficients}
a_s^\pm(T) &=& (2-\alpha)\  z_0^{1-\alpha}\ c_0^{\pm}\ t |t|^{-\alpha} \; , \nonumber \\
p_s^-(T) &=& (2-\alpha-\beta\delta)\ (-z_0 t)^{1-\alpha- \beta\delta} \; ,  \\
p_s^+(T) &=& (2-\alpha- 2 \beta\delta)\ c_1^+ (z_0 t)^{1-\alpha-2 \beta\delta} \nonumber
\;.
\end{eqnarray}\end{linenomath*}
Again we follow the notation of Ref.~\cite{Engels:2011km} with $c_0^\pm$,
$c_1^+$, and $a_1$ denoting coefficients appearing in the 
parameterization of the scaling function $f_f$.

In the chiral limit, $F_q$ and its temperature derivative become 
\begin{linenomath*}\begin{equation}
        \frac{F_q(T,0)}{T} = a^r_{0,0} + t \left( a^r_{1,0} + A^\pm |t|^{-\alpha} \right) \; ,
\label{FqH0}
\end{equation}\end{linenomath*}
with $A^\pm=(2-\alpha) z_0^{1-\alpha} c_0^\pm A$, and 
\begin{linenomath*}\begin{equation}
    T_c \frac{\partial (F_q(T,0)/T)}{\partial T} =  a^r_{1,0} \left( 1 + R^\pm  
       |t|^{-\alpha} \right)  \; ,
\label{FqH02}
\end{equation}\end{linenomath*}
with $R^\pm=(1-\alpha)A^\pm/a^r_{1,0}$.
The temperature derivative of an observable scaling as an energy density is
expected to scale as a specific heat; correspondingly we will call these
observables $C_V$-like, and we expect eq.~\eqref{FqH02} to exhibit $C_V$-like
characteristics.  
 

Similarly the baryon chemical potential $\mu_B$ is an energy-like coupling. 
We can include $\mu_B$ along with temperature in a general energy-like coupling,
which we also label $t$:
\begin{equation}\label{eq:genEnergyLike}
t\equiv\frac{1}{t_0}\left(\frac{T-T_c}{T_c}+\kappa \muh_B^2\right),
\end{equation}
where $t_0$ and $\kappa$ are non-universal constants. From
eq.~\eqref{eq:genEnergyLike} we see that taking two $\muh_B$ derivatives is the
same as taking one $T$ derivative, which along with eq.~\eqref{eq:fresponse}
leads to the expectation that $\fresponse$
should also be $C_V$-like. Taking derivatives of eq.~\eqref{Fqcritical}
and utilizing eq.~\eqref{eq:genEnergyLike}, one expects
\begin{equation}\label{eq:chiScaling}
  \fresponse = 2\kappa z_0A H^{-\alpha/\beta\delta}f_f''(z)
                                    +\frac{\partial^2\freg}{\partial \muh_B^2}.
\end{equation}
Comparing eq.~\eqref{eq:chiScaling} with eq.~\eqref{eq:FqT}, one sees clearly
that the observables should behave qualitatively similarly in the chiral limit.

%
%
\section{Computational setup and observables}

Our study uses $N_f=2+1$ HISQ configurations
with $m_s$ at its physical value and $m_l$ varying in the range
$H=m_l/m_s=1/160-1/20$. These include gauge field ensembles 
generated previously by the HotQCD collaboration
\cite{Bazavov:2011nk,Bazavov:2014pvz,bazavov_qcd_2017,HotQCD:2019xnw}
and some further configurations for $H=1/40$ and $H=1/80$.
The bare coupling $\beta$ for $H\leq1/27$ is taken in the range 
6.26-6.50, which is chosen so $T$ is near the chiral pseudocritical
temperature. For $H=1/20$ we also use data from calculations on lattices
at smaller couplings, $\beta=6.05, 6.125$, and $6.175$
\cite{Bazavov:2013yv}, which helps establish contact to the low
$T$ regime. 
We set the scale with $f_K$~\cite{Bazavov:2010hj} using a recent
parameterization of lattice QCD results for $f_K\,a(\beta)$
\cite{Bazavov:2019www}.
We find no significant dependence of $\ev{P}$ on
spatial volume~\cite{Clarke:2020clx} and hence, we use the gauge ensembles 
with the largest volumes in our analysis.
A summary of statistics used to analyze the scaling behavior of $\ev{P}$, $F_q$,
and their derivatives is given in 
Ref.~\cite{Clarke:2020clx}.

The Polyakov loop requires a multiplicative renormalization,
\begin{linenomath*}\begin{equation}
  P = e^{-c(g^2) N_\tau}\Pbare,
\end{equation}\end{linenomath*}
{\it i.e.} the renormalized $P$ appears in eq.~\eqref{eq:Fav}.
When needed, renormalization constants $c(g^2)$
are obtained from an interpolation of Table V of
Ref.~\cite{Bazavov:2016uvm}. 
We note however that derivatives of the free energy such as
\begin{linenomath*}\begin{equation}
  \frac{\partial F_q/T}{\partial H} 
  = -\frac{1}{\ev{P}} \frac{\partial \ev{P}}{\partial H}
  = \ev{P\cdot \Psi}-\ev{P}\ev{\Psi}
~~~~~\text{and}~~~~~
  T_c \frac{\partial F_q/T}{\partial T} 
  = -\frac{T_c}{\ev{P}} \frac{\partial \ev{P} }{\partial T}
  \label{eq:FqHandFqT}
\end{equation}\end{linenomath*}
are independent of the renormalization in the continuum limit,
since this cancels out in the ratio. Here $\Psi\equiv\frac{1}{2}\hat{m_s}\tr D_l^{-1}$
is an extensive observable written in terms of the dimensionless bare strange quark
mass $\hat{m_s}$ and the staggered Dirac matrix $D_l$.
Similarly $\fresponse$, which can be extracted
as~\cite{delia_dependence_2019}
\begin{equation}
  \fresponse=\frac{1}{9}\left( \frac{\ev{\Re P\;(n^2+n')}}{\ev{\Re P}} - \ev{n^2+n'}
             +\frac{\ev{(\Re P+\Im P)\;n}^2}{\ev{\Re P}^2} \right),
\end{equation}
where the total quark number $n$ in the $N_f=2+1$ HISQ formulation is
\begin{equation}
  n=2n_l+n_s,~~~~~~~~~~~n_f=\frac{1}{4}\tr D_f^{-1}\partial_{\muh_f}D_f, 
             ~~~~~~~~~~~n'=2\partial_{\muh_l}n_l+\partial_{\muh_s}n_s,
\end{equation}
should be renormalization independent. To extract $\fresponse$ we used about 
2.5 million molecular dynamic time units (MDTU) per temperature for $H=1/27$ 
and about 125,000 MDTU for $H=1/40$, 
including some newly generated configurations for the latter.

%
%
\section{Results}

\begin{figure}
\centering
\includegraphics[width=0.47\textwidth]{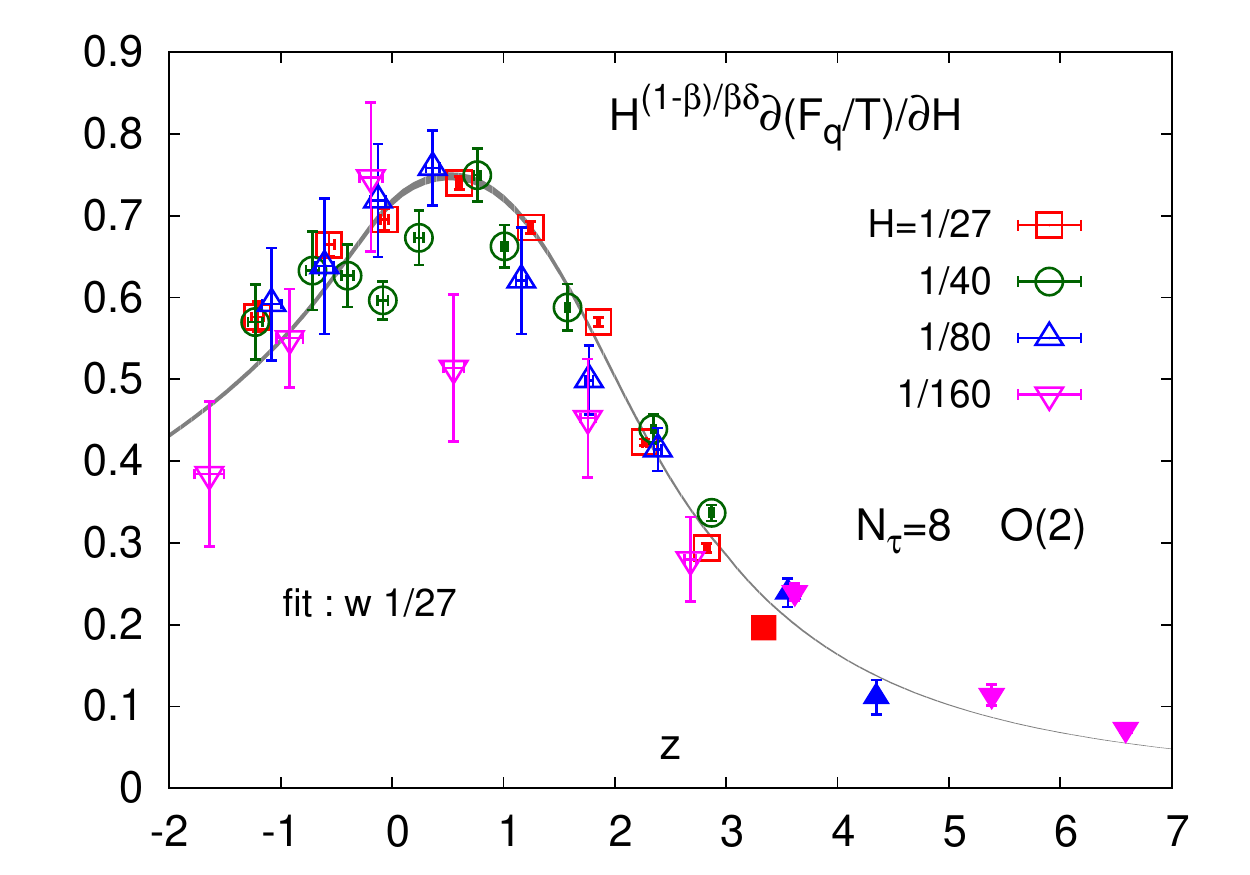}
\includegraphics[width=0.47\textwidth]{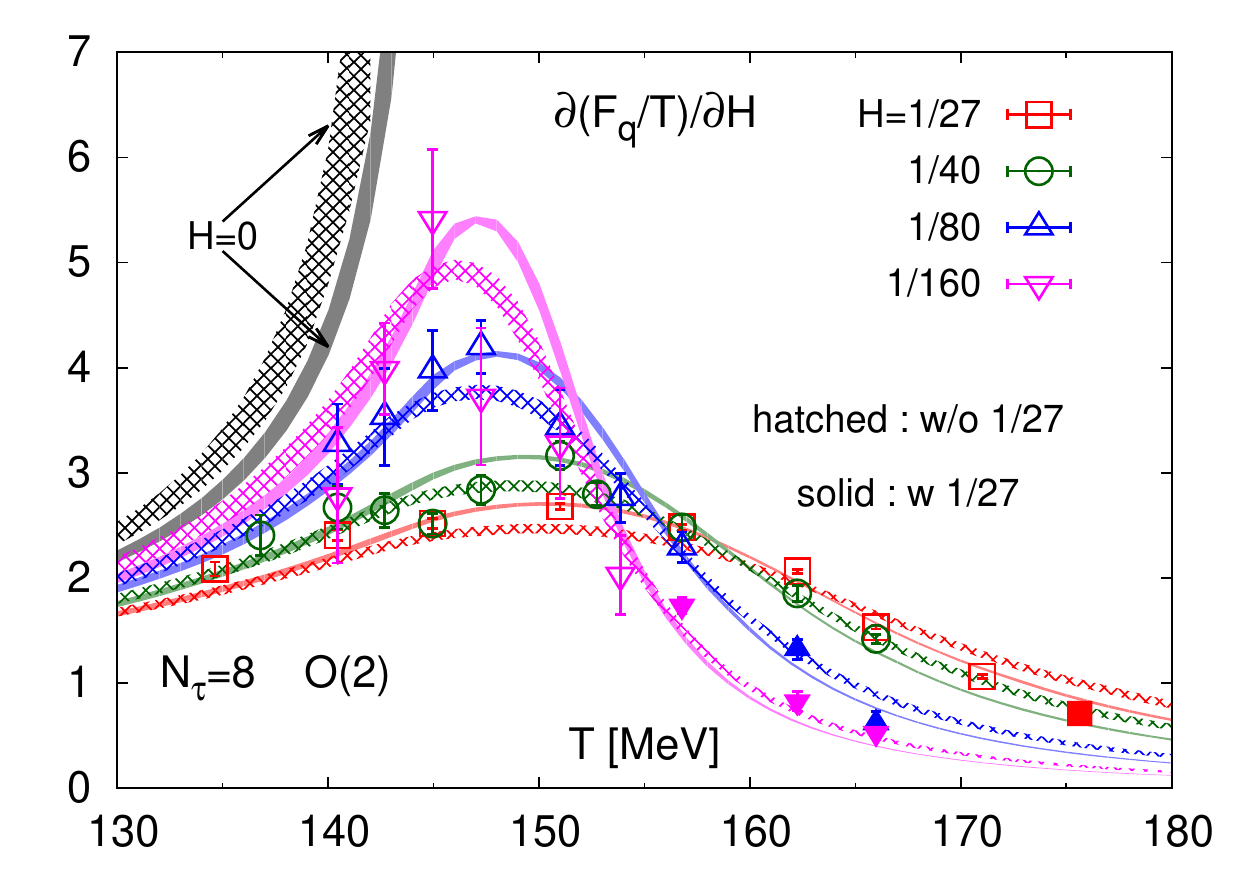}
\caption{Scaling and $T$-dependence of $\partial (F_q/T)\partial H$.
         Data with filled symbols are not included in the fits.
         {\it Left}: Rescaled $\partial (F_q/T)/\partial H$ 
         as a function of $z$. {\it Right}: Derivative of $\partial
         (F_q/T)/\partial H$ as a function of $T$. 
         The chiral limit result obtained from this fit is shown as 
         grey bands. Solid curves are fits that include $H=1/27$ data, 
         while hatched curves are fits that exclude this these points.
         Images taken from Ref.~\cite{Clarke:2020htu}.}
\label{fig:dFqdHfit}
\end{figure}

In Fig.~\ref{fig:dFqdHfit} (left) we show data for $\partial (F_q/T)/\partial H$
rescaled with the power of $H$ according eq.~\eqref{Fqm-crit}. 
That data for different $H$ largely fall on top of each other
suggests that $\partial (F_q/T)/\partial H$ indeed diverges 
as $H^{(\beta-1)/\beta\delta}$ as $H\to0$ and that $H$-dependent
contributions to $F_q/T$ originating from $\freg$ are 
small compared to those coming from the singular part. We therefore 
employ an $H$-independent fit ansatz for $\freg$,
{\it i.e.} we use $\freg (T,H=0)$ in every fit.
We performed this 3-parameter fit for each data set by either
including or leaving out the $H=1/27$ data; the results are shown in
Fig.~\ref{fig:dFqdHfit}~(right). The fit parameters
$A$, $T_c$, and $z_0$ are given in Table I of Ref. \cite{Clarke:2020htu}.
The parameters $T_c$ and $z_0$ agree well with earlier results 
for chiral susceptibilities in (2+1)-flavor QCD \cite{HotQCD:2019xnw}.

\begin{figure}
\centering
\includegraphics[width=0.47\textwidth]{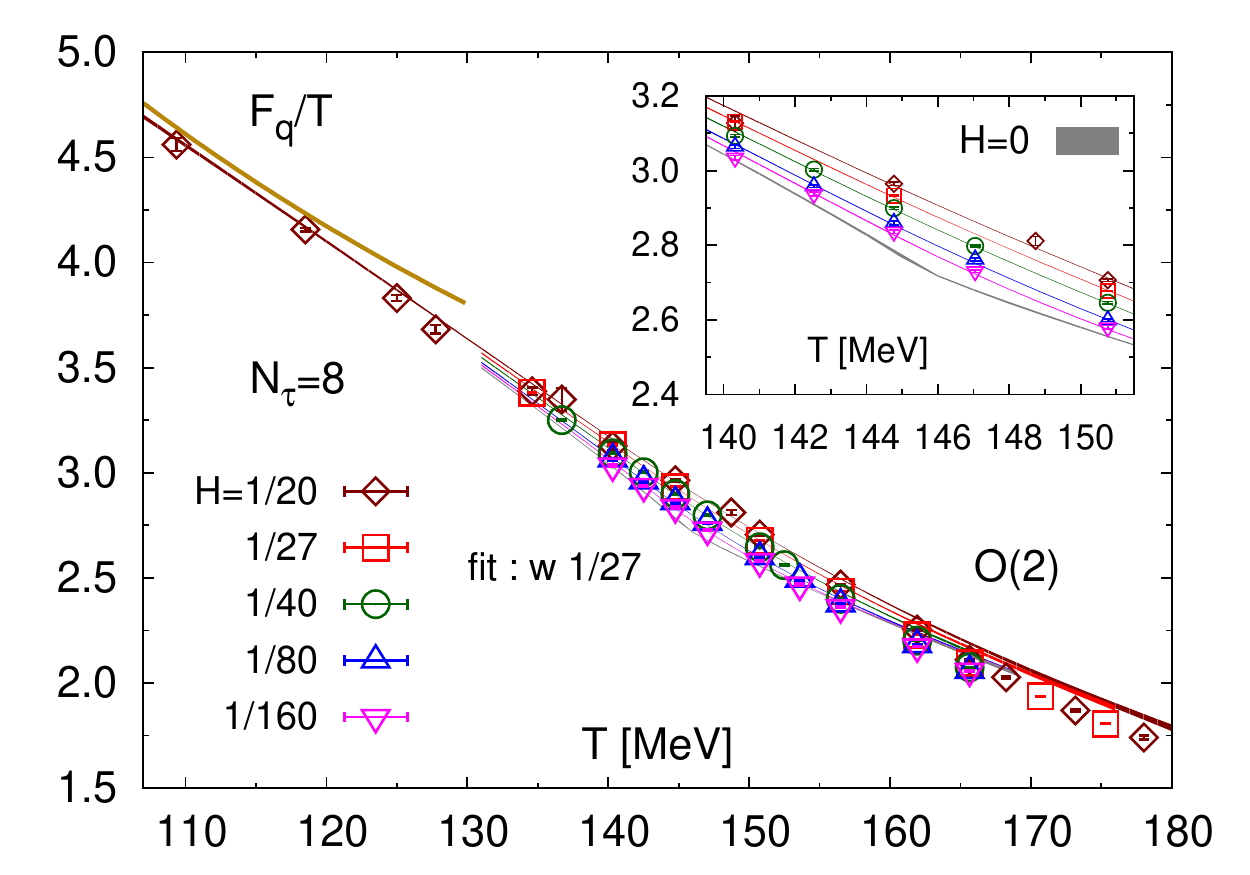}
\includegraphics[width=0.47\textwidth]{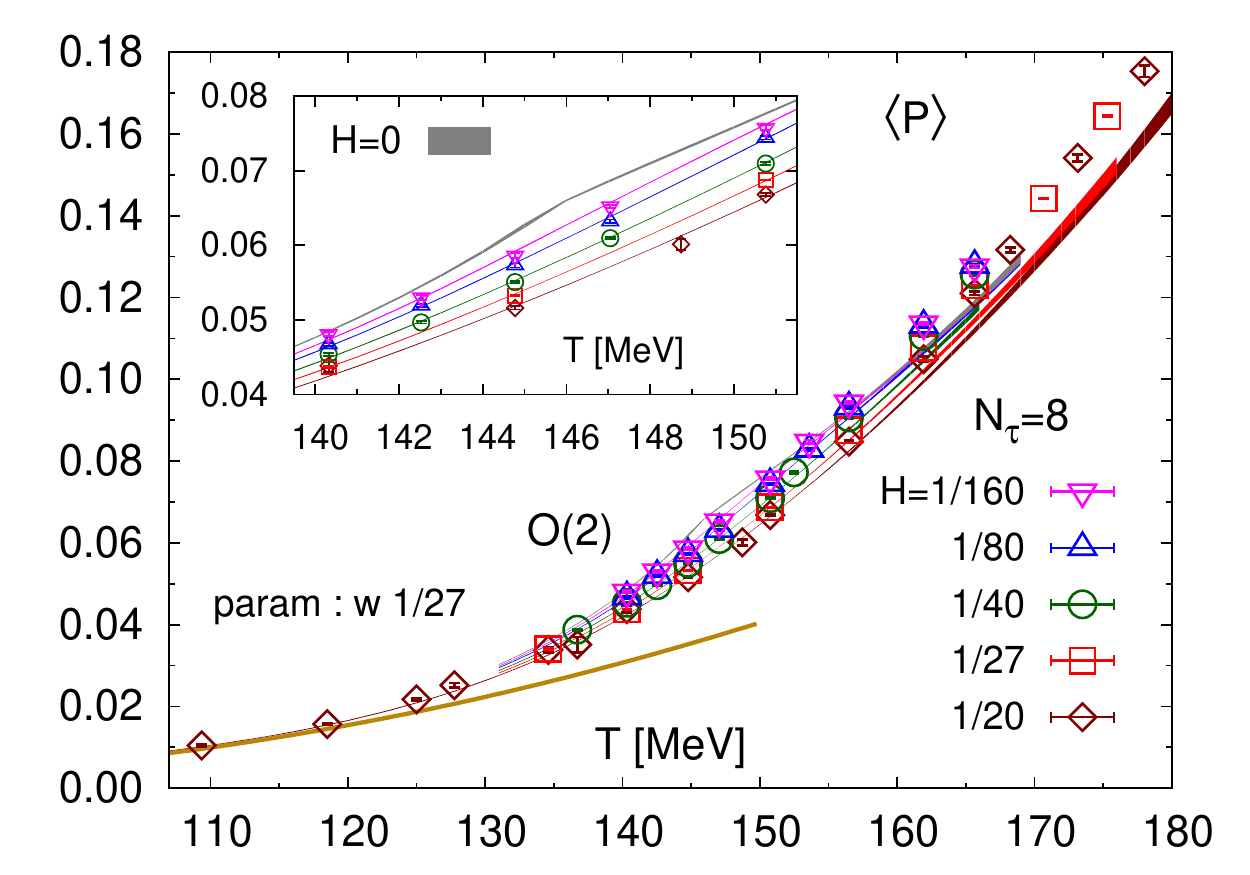}
\caption{Chiral fits for $F_q/T$ and $\ev{P}$. The insets show data in the $T$ 
         range covered by the fit. Both fits include $H=1/27$ data.
         The chiral limit results obtained from the fit are shown as a grey band.
         The solid gold line shows the heavy-light
         meson contribution calculated in the hadron-gas approximation
         \cite{Bazavov:2013yv}. {\it Left}: $T$-dependence of $F_q/T$.
         {\it Right}: $T$-dependence of $\ev{P}$. Images taken from Ref.
         \cite{Clarke:2020htu}.} 
\label{fig:Fqfit}
\end{figure}

Looking back at eq.~\eqref{O4criticalmass}, it is instructive to note that below $T_c$ 
the next most significant term in $H$, which goes as $H^{3/2}$, will also
receive contributions from singular terms only. This
term turns out to be negative; therefore one expects
$\partial(F_q/T)/\partial H$ to increase with decreasing $H$ toward
the chiral limit behavior, $Ap_s^-(T)$. 
From the hyperscaling relation
$\alpha=2-\beta(1+\delta)$, this is proportional to
$|t|^{\beta-1}$.
Correspondingly in Fig.~\ref{fig:Fqfit} (right) one sees an increase with decreasing $H$
below $T_c$ toward the expected chiral limit $T$-dependence.

Next, following our master equation~\eqref{Fqcritical} we assume for $F_q/T$ a functional form
\begin{equation}
\frac{F_q}{T}
  \approx{A}H^{(1-\alpha)/\beta\delta}\,
 f'_f\left({z_0}\frac{T-T_c}{T_c}H^{-1/\beta\delta}\right)
 + {a^r_{0,0}} + a^r_{1,0}t
\end{equation}
keeping the leading $t$-dependence in the regular term. We use the
previously determined $A$, $T_c$, and $z_0$ and fit the remaining two regular
parameters to the $F_q/T$ data, shown in Fig.~\ref{fig:Fqfit} (left).
The $T$ range and data
used in the fit are shown in the inset. The  resulting fit
parameters $a^r_{0,0}$ and $a^r_{1,0}$ are also given in Table I of Ref.
\cite{Clarke:2020htu}. We also show in this figure the
heavy-light meson contribution to $F_q/T$ calculated in the hadron-gas
approximation \cite{Megias:2012kb,Bazavov:2013yv}.

Once we have determined all five fit parameters for $F_q/T$, we can
plug them into eq.~\eqref{Pcritical} to arrive at 
the $T$ and $H$ dependence of $\ev{P}$. The thus
determined curves are shown in Fig.~\ref{fig:Fqfit} (right).
As seen in the inset, they agree well with $\ev{P}$
data near $T_c^{N_\tau=8}=144(2)$~MeV \cite{HotQCD:2019xnw},
which suggests the behavior of $\ev{P}$ is explained
well by chiral scaling in this region and serves
as a consistency check of our approach.

\begin{figure}
\centering
\includegraphics[width=0.45\textwidth]{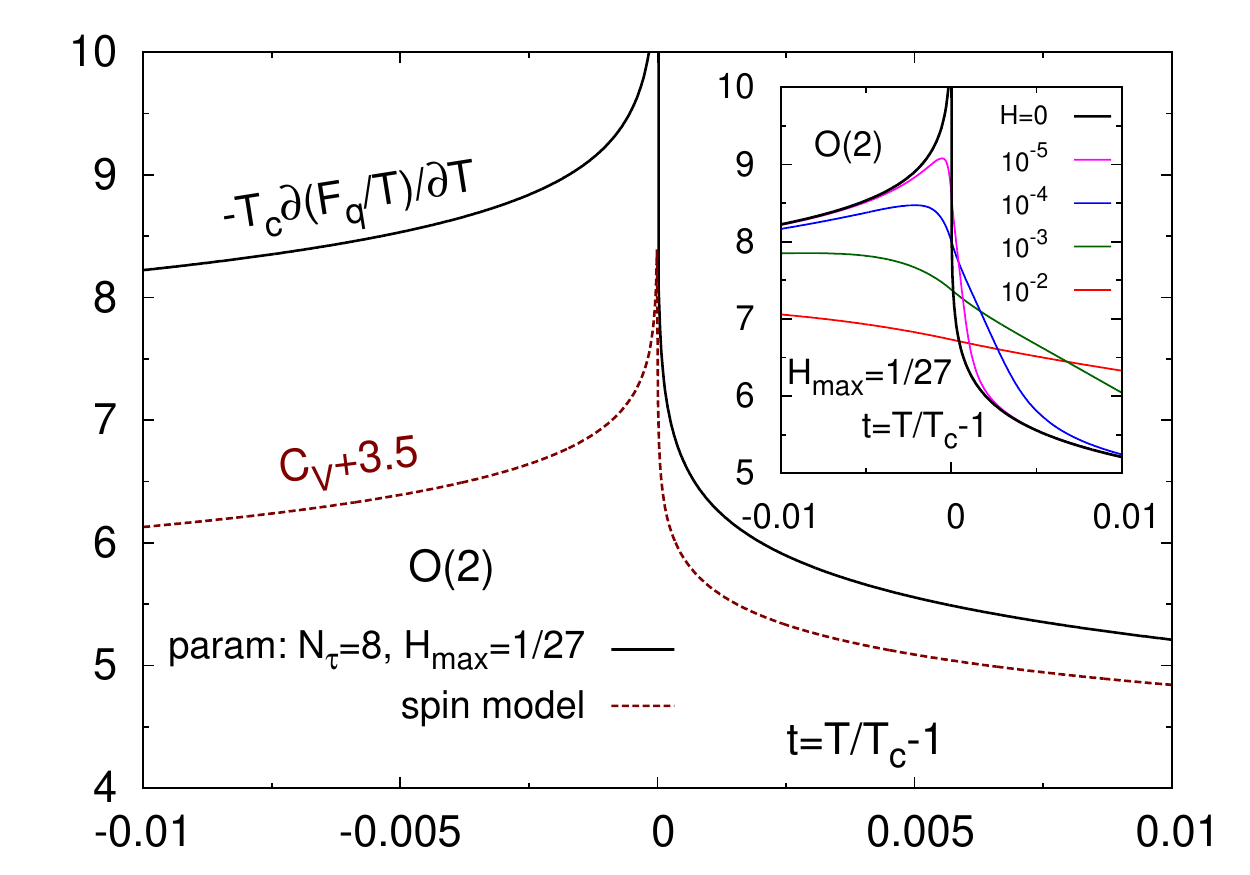}
\includegraphics[width=0.45\textwidth]{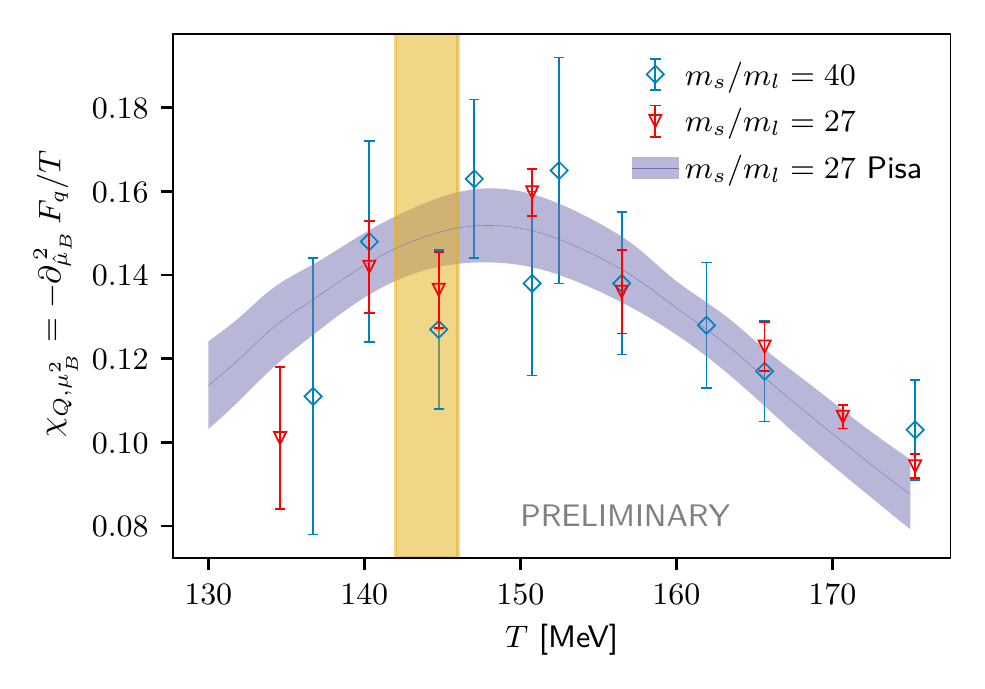}
\caption{{\it Left:} Comparison of $C_V$ at $H=0$ for the 3-$d$, $\O(2)$
         spin model taken from Ref.~\cite{Cucchieri:2002hu}
         (dashed line) and $-T_c \partial(F_q(T,0)/T)/\partial T$ (solid line). The 
         former curve is shifted vertically by a constant 
         for easier comparison. The inset shows how the spike in
         $-T_c \partial(F_q/T)/\partial T$ develops in the chiral limit.
         Image taken from Ref.~\cite{Clarke:2020htu}.
         {\it Right:} Response of $F_q/T$ to $\mu_B$ as a function of 
         $T$. The vertical yellow band indicates $T_c^{N_\tau=8}$
         \cite{HotQCD:2019xnw}. The blue band is an interpolation based on results
         from Ref.~\cite{delia_dependence_2019}.}
\label{fig:CVspike}
\end{figure}

In the chiral limit, the $T$-dependence of both $F_q/T$ and $\ev{P}$ exhibit a
small kink at $T_c$. This kink hints toward a characteristic spike found in
$C_V$-like observables in $\O(N)$ universality classes, which we examine using
eqs.~\eqref{FqH0} and \eqref{FqH02}.
Although at $T_c$ the contribution to the slope of $F_q/T$ is entirely given by
the regular term $a_{1,0}^r$, near $T_c$ this
contribution gets mostly canceled by
the singular contributions $A^\pm |t|^{-\alpha}$. 
This is the origin of the spike. In the chiral limit our fit
results suggest the appearance of this spike in the
$T$-derivatives of $F_q(T,0)/T$ as well as
$\ev{P}$.

The basic features found in our analysis of $F_q/T$ and $\ev{P}$ 
are quite similar to those found in the analysis of 3-$d$, $\O(2)$ symmetric spin models 
\cite{Cucchieri:2002hu}. Also in that case a large 
cancellation of contributions arising from regular
and singular terms is found; the spike in $C_V$ 
is concentrated in a temperature interval 
of about 1\% around $T_c$, where $C_V$ changes by almost 
a factor $10$. In Fig.~\ref{fig:CVspike} (left) we show a comparison of the
behavior of $C_V$ in $\O(2)$ spin models \cite{Cucchieri:2002hu} and
our result for $-T_c\,\partial (F_q/T)/\partial T$ at $H=0$.
The inset shows the development of this sharp peak as $m_l$
decreases. Clearly, this feature becomes visible only for
$H$ being substantially smaller than the light quark masses 
$H\approx 10^{-2}$ that are
accessible in modern lattice QCD calculations. 

Our preliminary results for the susceptibility $\fresponse$ 
at $H=1/27$ and $H=1/40$ are shown in Fig.~\ref{fig:CVspike} 
(right) along with an interpolation of $H=1/27$ data from 
Ref.~\cite{delia_dependence_2019}. Here we find excellent agreement
with their results at physical $m_l$. At our current statistical power, 
our results for $H=1/40$ are consistent with $H=1/27$.
Besides the fact that $\fresponse$ is somewhat noisy, the ability to
resolve a clear peak will depend on the relative importance of singular
and regular contributions\footnote{We encounter this problem with
other observables expected to be $C_V$-like. For example fourth-order
conserved charge fluctuations are $C_V$-like, and for this reason
one generally expects
to find a peak near $T_c$; nevertheless contributions
of the regular terms to the fourth-order strangeness fluctuation are
significant enough to render it monotonic near $T_c$~\cite{Sarkar:2020soa}.
Similarly the Polyakov loop susceptibility is $C_V$-like. At finite $N_\tau$
one finds monotonic behavior near $T_c$~\cite{Clarke:2019tzf}, which may again
be due to significant regular contributions.}.
To try to address these issues, we plan to increase the
statistics and lower $H$, where we expect the apparent peak in this
quantity to shift closer to $T_c$. 
We hope to eventually carry out a complete
scaling analysis as has been done with the other $C_V$-like observables.

\section{Conclusion and outlook}

For $N_f=2+1$ HISQ fermions at $N_\tau=8$ near and below physical $m_l$, 
we find $\partial_H F_q$ diverges as $H\to 0$ according to the 3-$d$, 
$\O(2)$ universality class. The Polyakov loop is described well by the 
3-$d$, $\O(2)$ scaling function near $T_c$, and $\partial_T F_q$ behaves 
qualitatively similarly as $C_V$ in $\O(2)$ spin models. We expect this 
behavior to be consistent with $\O(4)$ in the continuum limit. We stress that
the lack of an {\it a priori} reason to associate $\ev{P}$ with deconfinement
along with its consistency with $\O(N)$ scaling cast serious doubt on
attempts to interpret it as an indicator for the deconfinement
of light degrees of freedom near and below physical $m_l$.

We are in the process of investigating the chiral behavior of 
$\fresponse$, which we expect also to be $C_V$-like.
More statistics are required at $H=1/40$. 
We plan to calculate this observable for $H=1/80$ and
smaller in order to also investigate its scaling behavior.

\section*{Acknowledgements}

This work was supported by the Deutsche Forschungsgemeinschaft (DFG,
German Research Foundation) Proj. No. 315477589-TRR 211; and by the
German Bundesministerium f\"ur Bildung und Forschung through Grant No.
05P18PBCA1. We thank HotQCD for providing access to their latest data
sets and for many fruitful discussions.

\bibliographystyle{JHEP}
\bibliography{bibliography}

\end{document}